\documentclass[12pt]{iopart}

\usepackage{cite}

\usepackage{graphicx}

\usepackage{dcolumn}

\begin{document}

\title{The virial theorem for nonlinear problems}

\author{Paolo Amore\dag \ and Francisco M Fern\'andez
\footnote[2]{Corresponding author}}

\address{\dag\ Facultad de Ciencias, Universidad de Colima, Bernal D\'iaz del
Castillo 340, Colima, Colima, Mexico}\ead{paolo.amore@gmail.com}

\address{\ddag\ INIFTA (UNLP, CCT La Plata-CONICET), Divisi\'on Qu\'imica Te\'orica,
Blvd. 113 S/N,  Sucursal 4, Casilla de Correo 16,
1900 La Plata, Argentina}\ead{fernande@quimica.unlp.edu.ar}

\maketitle

\begin{abstract}
We show that the virial theorem provides a useful simple tool for
approximating nonlinear problems. In particular we consider conservative
nonlinear oscillators and a bifurcation problem. In the former case we
obtain the same main result derived earlier from the expansion in Chebyshev
polynomials.
\end{abstract}

\section{Introduction \label{sec:Intro}}

In a recent paper Bel\'{e}ndez et al\cite{BAFP09} showed that the widely
used small--amplitude approximation cannot always be successfully applied to
nonlinear oscillators. To overcome this difficulty the authors proposed the
expansion of the nonlinear force in terms of Chebyshev polynomials. This
alternative linearization of nonlinear problems proved to be remarkably more
accurate and efficient than the straightforward small--amplitude approach.
Besides, the Chebyshev series applies even to such difficult cases where the
Taylor series fails\cite{BAFP09}.

The purpose of present article is to discuss an alternative approach to
nonlinear problems based on the well--known virial theorem\cite{G80}. In Sec~%
\ref{sec:oscillators} we outline the main results of Bel\'{e}ndez et al\cite
{BAFP09} for conservative nonlinear oscillators. In Sec.~\ref{sec:VT} we
develop the virial theorem, apply it to conservative nonlinear oscillators,
and compare its results with those obtained by Bel\'{e}ndez et al\cite
{BAFP09}. In Sec.~\ref{sec:bifurcation} we apply the virial theorem to a
nonlinear problem that exhibits bifurcation and compare its results with the
exact ones and with those produced by the small--amplitude approximation. In
Sec.~\ref{sec:conclusions} we discuss the main results of the paper and draw
conclusions.

\section{Conservative nonlinear oscillators \label{sec:oscillators}}

Bel\'{e}ndez et al\cite{BAFP09} considered nonlinear conservative autonomous
systems given by the second--order differential equation
\begin{equation}
\ddot{x}+f(x)=0  \label{eq:dif_eq}
\end{equation}
with the boundary conditions $x(0)=A$, $\dot{x}(0)=0$. Here a point
indicates derivative with respect to $t$. In particular, Bel\'{e}ndez et al%
\cite{BAFP09} restricted themselves to odd functions $f(-x)=-f(x)$ that
satisfy $xf>0$.

The approach proposed by Bel\'{e}ndez et al\cite{BAFP09} consists in the
expansion of the force in a series of Chebyshev polynomials of the first
kind $T_{n}(z)$:
\begin{equation}
f(x)=\sum_{n=0}^{\infty }b_{2n+1}(A)T_{2n+1}(y)  \label{eq:f_Cheby}
\end{equation}
where $y=x/A$. These polynomials are given by the recurrence relation
\begin{eqnarray}
T_{0}(z) &=&1  \nonumber \\
T_{1}(z) &=&z  \nonumber \\
T_{n+1}(z) &=&2zT_{n}(z)-T_{n-1}(z)  \label{eq:Cheby_rec}
\end{eqnarray}
and are orthogonal in $-1\leq z\leq 1$ with the weight function $%
w(z)=(1-z^{2})^{-1/2}$:
\begin{equation}
\int_{-1}^{1}(1-z^{2})^{-1/2}T_{m}(z)T_{n}(z)=\frac{\pi }{2}(1+\delta
_{m0})\delta _{mn}  \label{eq:orthogonal}
\end{equation}
Therefore, the coefficients of the expansion (\ref{eq:f_Cheby}) are given by
\begin{equation}
b_{2n+1}(A)=\frac{2}{\pi }\int_{-1}^{1}(1-y^{2})^{-1/2}T_{2n+1}(y)f(Ay)\,dy
\label{eq:bn}
\end{equation}
Notice that there is a misprint in the weight function shown by Bel\'{e}ndez
et al\cite{BAFP09}.

If we keep only the first term in the expansion (\ref{eq:f_Cheby}) the
differential equation (\ref{eq:dif_eq}) becomes that for a harmonic
oscillator
\begin{equation}
\ddot{x}+\frac{b_{1}(A)}{A}x=0  \label{eq:dif_eq_lin}
\end{equation}
with a frequency
\begin{equation}
\omega =\sqrt{\frac{b_{1}(A)}{A}}  \label{eq:omega_app}
\end{equation}
that depends on the amplitude $A$. This expression proves to be remarkably
accurate for many problems\cite{BAFP09} in spite of its simplicity.

\section{The virial theorem \label{sec:VT}}

Here we consider the same differential equation (\ref{eq:dif_eq}) with the
more general boundary conditions
\begin{equation}
x(b)\dot{x}(b)-x(a)\dot{x}(a)=0  \label{eq:BC_gen}
\end{equation}
If we integrate the equation
\begin{equation}
\frac{d}{dt}x^{n}\dot{x}=nx^{n-1}\dot{x}^{2}+x^{n}\ddot{x}=nx^{n-1}\dot{x}%
^{2}-x^{n}f
\end{equation}
we obtain
\begin{equation}
n\int_{a}^{b}x^{n-1}\dot{x}^{2}\,dt=\int_{a}^{b}x^{n}f\,dt+x(b)^{n}\dot{x}%
(b)-x(a)^{n}\dot{x}(a)  \label{eq:HT}
\end{equation}
In particular, when $n=1$ we have
\begin{equation}
\int_{a}^{b}\dot{x}^{2}\,dt=\int_{a}^{b}xf\,dt  \label{eq:VT}
\end{equation}
because of the boundary conditions (\ref{eq:BC_gen}).

We now apply this general expression to the oscillators studied by
Bel\'{e}ndez et al\cite{BAFP09} that are periodic of period $\tau $. In this
case the kinetic energy is
\begin{equation}
K=\frac{\dot{x}^{2}}{2}  \label{eq:K}
\end{equation}
and if we choose $a=0$ and $b=\tau $ equation (\ref{eq:VT}) becomes the
well--known virial theorem\cite{G80}
\begin{equation}
2\bar{K}=\overline{xf}  \label{eq:VT_part}
\end{equation}
where the expectation values are defined as
\begin{equation}
\bar{F}=\frac{1}{\tau }\int_{0}^{\tau }F\,dt  \label{eq:exp_val}
\end{equation}
The virial theorem is known from long ago\cite{G80}; its name comes from the
fact that $xf$ is known as the virial of the forces in the mechanical
system. This theorem reveals the balance between the kinetic and potential
energies\cite{G80}.

The exact trajectory $x(t)$ satisfies equation (\ref{eq:VT_part}). If we
propose an approximate trajectory of the form
\begin{equation}
x_{app}(t)=A\cos (\omega t)  \label{eq:x_app}
\end{equation}
where $\omega =2\pi /\tau $ is the frequency of the oscillator, then it is
reasonable to set this approximate frequency so that $x_{app}(t)$ satisfies
the virial theorem (\ref{eq:VT_part}). If we substitute equation (\ref
{eq:x_app}) into equation (\ref{eq:VT_part}) we obtain
\begin{equation}
\pi \omega A^{2}=2\int_{0}^{\tau /2}xf\,dt=\frac{2A}{\omega }\int_{-1}^{1}%
\frac{yf(Ay)}{\sqrt{1-y^{2}}}\,dy
\end{equation}
by means of the change of variables $y=\cos (\omega t)$. This is exactly the
equation for the frequency (\ref{eq:omega_app}) derived by Bel\'{e}ndez et al%
\cite{BAFP09}.

We appreciate that both the virial theorem and the first term of the
Chebyshev expansion lead to the same approximate frequency.

\section{Bifurcation \label{sec:bifurcation}}

Equation (\ref{eq:VT}) is sufficiently general for the treatment of a wide
variety of interesting nonlinear problems of the form (\ref{eq:dif_eq}). In
this section we consider the Bratu equation
\begin{equation}
u^{\prime \prime }(x)+\lambda e^{u(x)}=0,\,u(0)=u(1)=0  \label{eq:Bratu}
\end{equation}
that appears in simple models for the study spontaneous explosion due to
internal heating in combustible materials\cite{VA93,MO05}. It is also
interesting for another reason: it is a simple strongly nonlinear problem
that can be exactly solved. Therefore, it is not surprising that it has
become a useful benchmark for testing approximate methods\cite
{MO05,W89,M98,B03,B04}.

It is well--known that the solution to the Bratu equation is\cite{B03}
\begin{equation}
u(x)=-2\ln \left\{ \frac{\cosh \left[ \theta (x-1/2)\right] }{\cosh (\theta
/2)}\right\}  \label{eq:Bratu_exact}
\end{equation}
where $\theta $ is a root of
\begin{equation}
\lambda =\frac{2\theta ^{2}}{\cosh (\theta /2)^{2}}  \label{eq:lambda(theta)}
\end{equation}
This equation exhibits two solutions when $\lambda <\lambda _{c}$, only one
when $\lambda =\lambda _{c}$, and none when $\lambda >\lambda _{c}$, where
the critical $\lambda $--value $\lambda _{c}$ is the maximum of $\lambda
(\theta )$. We easily obtain it from the root of $d\lambda (\theta )/d\theta
=0$ that is given by
\begin{equation}
e^{\theta _{c}}(\theta _{c}-2)-\theta _{c}-2=0  \label{eq:theta_c}
\end{equation}
The exact critical parameters are $\theta _{c}=2.399357280$ and $\lambda
_{c}=3.513830719$.

\begin{figure}[H]
\begin{center}
\includegraphics[width=9cm]{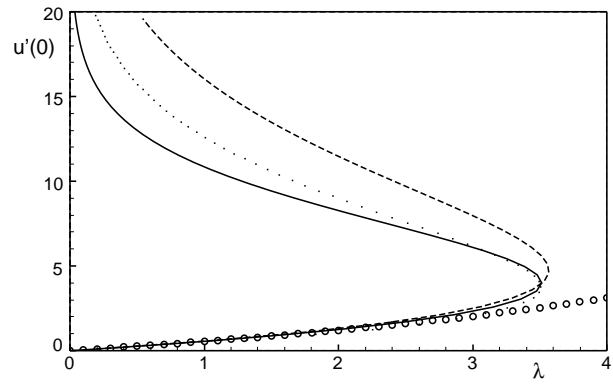}
\end{center}
\caption{Bifurcation diagram for the slope at origin $u^\prime (0)$ in terms
of $\lambda$ obtained by means of the exact expression (solid line), $%
u(x)=Ax(1-x)$ (dashed line), $u(x)=A\sin(\pi x)$ (dots) and Taylor
linearization (circles)}
\label{fig:Bratu}
\end{figure}

The slope at origin
\begin{equation}
u^{\prime }(0)=\frac{2\theta (e^{\theta }-1)}{(e^{\theta }+1)}
\label{eq:Bratu_u'(0)_exact}
\end{equation}
displays a bifurcation diagram as a function of $\lambda $ as shown in Fig.~%
\ref{fig:Bratu} (it is not difficult to obtain it by means of a parametric
plot using equations (\ref{eq:lambda(theta)}) and (\ref{eq:Bratu_u'(0)_exact}%
)). For the critical value of $\lambda $ we have $u^{\prime }(0)_{c}=4$.

In what follows we show that the virial theorem is suitable for estimating
the form of this bifurcation diagram. We simply have to introduce a trial
function $u(x)$, which satisfies the appropriate boundary conditions, into
the expression for the ``virial theorem''
\begin{equation}
\int_{0}^{1}u^{\prime 2}\,dx+\lambda \int_{0}^{1}ue^{u}\,dx=0
\label{eq:Bratu_VT}
\end{equation}
Notice that the exact solution satisfies $u^{\prime \prime }(x)<0$ for all $%
0<x<1$; therefore $u(x)$ is positive and do not have zeros between the end
points. This conclusion will guide us towards the choice of the trial
function.

One of the simplest functions that meets the criteria just indicated is
\begin{equation}
u(x)=Ax(1-x)  \label{eq:Bratu_trial1}
\end{equation}
A straightforward calculation shows that
\begin{equation}
\lambda =\frac{4A^{5/2}}{3\left[ \sqrt{\pi }(A-2)e^{A/4}\mathop{\rm erf}%
\left( \sqrt{A}/2\right) +2\sqrt{A}\right] }  \label{eq:Bratu_lambda_var1}
\end{equation}
and the slope at origin is $u^{\prime }(0)=A$, so that we can easily plot $%
u^{\prime }(0)$ vs $\lambda $ parametrically. Fig.~\ref{fig:Bratu} shows
that this expression is suitable fo the lower branch (small $\lambda $) but
it is not so accurate for the upper one (large $\lambda $). However, it
provides a reasonable description of the bifurcation diagram and the
critical parameters $\lambda _{c}=3.569086042$ and $u^{\prime
}(0)_{c}=4.727715383$ are remarkably close to the exact ones.

Another simple variational function that meets the required criteria is
\begin{equation}
u(x)=A\sin (\pi x)  \label{eq:Bratu_trial_sin}
\end{equation}
that leads to
\begin{equation}
\lambda =\frac{A\pi ^{3}}{2\left\{ 2+\pi \left[ I_{1}(A)+L_{1}(A)\right]
\right\} }  \label{eq:Bratu_lambda_2}
\end{equation}
where $I_{\nu }(z)$ and $L_{\nu }(z)$ stand for the modified Bessel and
Struve functions\cite{AS72}, respectively. In this case $u^{\prime }(0)=\pi
A $ and Fig.~\ref{fig:Bratu} shows that this expression is slightly less
accurate than the preceding one for the lower branch and certainly more
accurate for the upper one. Besides, this trial function yields better
critical parameters: $\lambda _{c}=3.509329130$ and $u^{\prime
}(0)_{c}=3.756549365$.

The Bratu equation is also suitable for revealing the limitation of the
linearization by means of an expansion in a Taylor series. If we neglect the
nonlinear terms in the expansion: $e^{u}=1+u+\ldots $ then we can solve the
resulting differential equation exactly and obtain
\begin{equation}
u(x)=\cos \left( \sqrt{\lambda }x\right) +\tan \left( \frac{\sqrt{\lambda }}{%
2}\right) \sin \left( \sqrt{\lambda }x\right) -1  \label{eq:Bratu_u_expan}
\end{equation}
In this case the slope at origin is
\begin{equation}
u^{\prime }(0)=\sqrt{\lambda }\tan \left( \frac{\sqrt{\lambda }}{2}\right)
\label{eq:Bratu_slope_expan}
\end{equation}
Fig.~\ref{fig:Bratu} shows that this approach based on the Taylor expansion
is unable to reproduce the upper branch of the bifurcation diagram. The
explanation is quite simple: the solution for the lower branch is
considerably smaller than the one for the upper branch. Therefore, an
expansion based on small values of $u$ will necessarily produce the former
and fail on the latter. On the other hand, an expansion in appropriate
orthogonal polynomials (or the virial theorem) provides an acceptable
description of both branches of the bifurcation diagram.

\section{Conclusions \label{sec:conclusions}}

We have shown that the approach derived by Bel\'{e}ndez et al\cite{BAFP09}
from the first term of the expansion in Chebyshev polynomials can also be
obtained by means of the virial theorem. It is clear that we can introduce
the approximation in two different ways: as the first term of a systematic
numerical method or as the requirement posed by the virial theorem with a
direct physical interpretation. One or the other point of view (or perhaps
one after the other) may be useful for teaching an undergraduate course on
classical mechanics. One can easily derive and discuss the virial theorem
for mechanical problems and then generalize it for the treatment of
arbitrary ordinary nonlinear differential equations. One advantage of the
approach based on the virial theorem is that it is also suitable for the
treatment of quantum--mechanical problems as well\cite{FC87}.

The virial theorem provides us with a quite general expression that may be
useful in the study of many nonlinear problems. As an example we have shown
that the approach is suitable for the treatment of the well--known Bratu
equation that appears in simple models for heat combustion\cite
{VA93,MO05,W89,M98,B03,B04}. In this case we have been able to try two
different approximate solutions which may probably be more difficult if one
merely resorts to an expansion in orthogonal polynomials.

\end{document}